\RequirePackage{fix-cm}
\documentclass[smallextended]{svjour3}       
\smartqed  
\usepackage{graphicx}
\usepackage{amsmath}
\usepackage{url}
\usepackage[vlined,boxed,commentsnumbered]{algorithm2e}

\newcommand{\bmpt}[1]{\begin{minipage}[t]{#1\textwidth}}
\newcommand{\bmp}[1]{\begin{minipage}{#1\textwidth}}

\newcommand{\emp}{\end{minipage}}

\newcommand{\cN}{{\cal N}}

\newcommand{\cO}{{\cal O}}
\newcommand{\bv}{\begin{verbatim}}
\newcommand{\ev}{\end{verbatim}}

\newcommand{\ben}{\begin{enumerate}}
\newcommand{\een}{\end{enumerate}}
\newcommand{\be}{\begin{equation}}
\newcommand{\ee}{\end{equation}}
\newcommand{\bea}{\begin{eqnarray}}
\newcommand{\eea}{\end{eqnarray}}
\newcommand{\beaa}{\begin{eqnarray*}}
\newcommand{\eeaa}{\end{eqnarray*}}
\newcommand {\bd}{\begin{description}}
\newcommand {\ed}{\end{description}}
\newcommand{\bi}{\begin{itemize}}
\newcommand{\ei}{\end{itemize}}
\newcommand{\bc}{\begin{center}}
\newcommand{\ec}{\end{center}}
\newcommand {\bq}{\begin{quotation}}
\newcommand {\eq}{\end{quotation}} 
\newcommand {\bdo}{\begin{document}} 
\newcommand {\edo}{\end{document}} 
\newcommand{\eql}{\: = \:}
\newcommand{\dfl}{\: \equiv \:}
\newcommand{\onder}[2]{#1_{\mbox{\scriptsize #2}}}
\newcommand {\boven}[2]{#1^{\mbox{\scriptsize #2}}}
\newcommand{\argmin}[1]{\mathop{\rm argmin}\limits_{#1}}
\newcommand{\klein}[2]{#1_{\mbox{\tiny #2}}}
\newcommand{\doo}{\mbox{do}}
\newcommand{\fracd}[2]{{{\displaystyle #1} \over {\displaystyle #2}}}

\newcommand{\intarg}[2]{\!#2\,d#1}
\newcommand{\del}[3]{\left#1 #3 \right#2}
\newcommand{\av}[1]{\del{<}{>}{#1}}
\newcommand{\bra}[1]{\del{<}{|}{#1}}
\newcommand{\ket}[1]{\del{|}{>}{#1}}
\newcommand{\avmini}[1]{\del{<}{>_{\setminus i}}{#1}}
\newcommand{\avminij}[1]{\del{<}{>_{\setminus i,j}}{#1}}
\newcommand{\sminij}{s_{\setminus i,j}}
\newcommand{\ui}{\avmini{u_i^2}}
\newcommand{\hi}{\avmini{h_i}}
\newcommand{\br}[1]{\del{\{}{\}}{#1}}
\newcommand{\ar}[1]{\del{(}{)}{#1}}
\newcommand{\bl}[1]{\del{[}{]}{#1}}

\newcommand {\Hxx}{H^{(xx)}}
\newcommand {\Hxy}{H^{(xy)}}
\newcommand {\Hyx}{H^{(yx)}}
\newcommand {\Hyy}{H^{(yy)}}
\newcommand {\Cxx}{C^{(xx)}}
\newcommand {\Cxy}{C^{(xy)}}
\newcommand {\Cyx}{C^{(yx)}}
\newcommand {\Cyy}{C^{(yy)}}
\newcommand {\cx}{c^{(x)}}
\newcommand {\hy}{h^{(y)}}
\newcommand{\bt}{\bar{t}}
\newcommand {\bx}{\bar{x}}
\newcommand {\bX}{\bar{X}}
\newcommand {\by}{\bar{y}}
\newcommand {\bw}{\bar{w}}
\newcommand {\bh}{\bar{h}}
\newcommand {\vs}{\vec{s}}
\newcommand {\vbfs}{{\bf s}}
\newcommand {\vv}{\vec{v}}
\newcommand {\vh}{\vec{h}}
\newcommand {\vk}{\vec{k}}
\newcommand {\vz}{\vec{z}}
\newcommand {\vx}{\vec{x}}
\newcommand {\valpha}{\vec{\alpha}}
\newcommand {\vm}{\vec{m}}
\newcommand {\vtheta}{\vec{\theta}}
\newcommand {\vgamma}{\vec{\gamma}}
\newcommand {\qd}{q_{\partial_i}}
\newcommand {\qm}{q_{\setminus i}}
\newcommand{\vxv}{\vec{x}_v}
\newcommand{\vxh}{\vec{x}_h}
\newcommand {\vy}{\vec{y}}
\newcommand {\bfy}{{\bf y}}
\newcommand {\bfx}{{\bf x}}
\newcommand {\bft}{{\bf t}}
\newcommand {\bfm}{{\bf m}}
\newcommand {\bfu}{{\bf u}}
\newcommand {\bfw}{{\bf w}}
\newcommand {\bfW}{{\bf W}}
\newcommand {\bfmu}{{\bf \mu}}
\newcommand {\vw}{\vec{w}}
\newcommand {\vb}{\vec{b}}
\newcommand {\vu}{\vec{u}}
\newcommand {\vW}{\vec{W}}
\newcommand {\vM}{\vec{M}}
\newcommand {\sign}{\mbox{sign}}
\newcommand {\sech}{\mbox{sech}}
\newcommand {\hmu}{h^{\mu}_j}
\newcommand {\gmu}{g^{\mu}_j}
\newcommand {\gjkmu}{g^{\mu}_{jk}}
\newcommand {\sia}{s_i^\alpha}
\newcommand {\sjb}{s_j^\beta}
\newcommand {\hia}{h_i^\alpha}
\newcommand {\hjb}{h_j^\beta}
\newcommand {\mia}{m_i^\alpha}
\newcommand {\mib}{m_i^\beta}
\newcommand {\mjb}{m_j^\beta}
\newcommand {\wijab}{w_{ij}^{\alpha\beta}}
\newcommand {\chiijab}{\chi_{ij}^{\alpha\beta}}
\newcommand {\Cijab}{C_{ij}^{\alpha\beta}}
\newcommand {\Hijab}{H_{ij}^{\alpha\beta}}
\newcommand {\he}{h_{\mbox{ext}}}
\newcommand {\xia}{x_{i,\alpha}}
\newcommand {\vxk}{\vec{x}_{\mbox{{\small known}}}}
\newcommand {\vxu}{\vec{x}_{\mbox{{\small unknown}}}}
\newcommand {\Pab}{P_{\alpha\beta}}
\newcommand {\dab}{\delta^{\alpha\beta}}
\newcommand {\dij}{\delta_{ij}}
\newcommand {\erf}{\mbox{Erf}}

\journalname{--}
\begin{document}
\title{The Variational Garrote}


\author{Hilbert J. Kappen   \and Vicen\c{c} G\'omez
}

\institute{Hilbert J. Kappen \at
        Donders Institute for Brain Cognition and Behaviour\\
        Radboud University Nijmegen\\
        6525 EZ Nijmegen, The Netherlands\\
              \email{b.kappen@science.ru.nl}           
           \and
        Vicen\c{c} G\'omez \at
        Donders Institute for Brain Cognition and Behaviour\\
        Radboud University Nijmegen\\
        6525 EZ Nijmegen, The Netherlands\\
              \email{v.gomez@science.ru.nl}           
}

\date{Received: date / Accepted: date}

\maketitle
\begin{abstract}

In this paper, we present a new variational method for sparse regression using
$L_0$ regularization. The variational parameters appear in the approximate
model in a way that is similar to Breiman's Garrote model.  We refer to this
method as the variational Garrote (VG).  We show that the combination of the
variational approximation and $L_0$ regularization has the effect of making the
problem effectively of maximal rank even when the number of samples is small
compared to the number of variables.  The VG is compared numerically with the
Lasso method, ridge regression and the recently introduced paired mean field
method (PMF) \cite{titsias2012}.  Numerical results show that
the VG and PMF yield more accurate predictions and more accurately reconstruct
the true model than the other methods. It is shown that the VG finds correct
solutions when the Lasso solution is inconsistent due to large input
correlations.  Globally, VG is significantly faster than PMF and tends to
perform better as the problems become denser and in problems with strongly
correlated inputs.  The naive implementation of the VG scales cubic with the
number of features.  By introducing Lagrange multipliers we obtain a dual
formulation of the problem that scales cubic in the number of samples, but
close to linear in the number of features.

\end{abstract}

\maketitle
\section{Introduction}
One of the most common problems in statistics is linear regression. Given 
$p$ samples of 
$n$-dimensional input data $x_i^\mu, i=1,\ldots,n$ and 1-dimensional output
data $y^\mu$, with $\mu=1,\ldots,p$, find weights $w_i, w_0$ that best
describe the relation
\begin{align}
\label{eq:regress}
y^\mu = \sum_{i=1}^n w_i x_i^\mu + w_0+\xi^\mu
\end{align}
for all $\mu$. $\xi^\mu$ is zero-mean noise with inverse variance $\beta$.

The ordinary least square (OLS) solution is given by $\vw = \chi^{-1} \vb$
and $w_0=\bar{y}-\sum_i w_i \bar{x}_i$, 
where $\chi$ is the input covariance matrix $\vb$ is the vector of
input-output covariances and $\bar{x}_i, \bar{y}$ are the mean values. 
There are several problems with the OLS approach. When $p$ is small, it
typically has a low prediction accuracy due to over fitting. In particular,
when $p<n$, $\chi$ is not of maximal rank and so its inverse is not uniquely defined. 
In addition, the OLS solution is not sparse: it will find a solution $w_i\ne 0$
for all $i$. Therefore, the interpretation of the OLS solution is often
difficult.

These problems are well-known, and there exist a number of approaches
to overcome these problems.
The simplest approach is called ridge regression. It adds a
regularization term $\frac{1}{2}\lambda \sum_i w_i^2$ with $\lambda>0$
to the OLS criterion. This has the effect that the input covariance matrix $\chi$ gets
replaced by $\chi +\lambda I$ which is of maximal rank for
all $p$. One optimizes $\lambda$ by cross validation. Ridge regression
improves the prediction accuracy but not the interpretability of the
solution.

Another approach is Lasso \cite{tibshirani96}. It solves the OLS
problem under the linear constraint $\sum_i |w_i|\le t$. This problem
is equivalent to adding an $L_1$ regularization term $\lambda \sum_i
|w_i|$ to the OLS criterion. The optimization of the quadratic error
under linear constraints can be solved efficiently. 
See \cite{friedman2009}
for a recent account.  Again, $\lambda$ or $t$ may be found through
cross validation.  The advantage of the $L_1$ regularization is that
the solution tends to be sparse. This improves both the prediction
accuracy and the interpretability of the solution.

The $L_1$ or $L_2$ regularization terms are known as shrinkage priors because
their effect is to shrink the size of $w_i$.  The idea of shrinkage prior has
been generalized by \cite{frank1993} to the form $\lambda \sum_i |w_i|^q$ with
$q>0$ and $q=1,2$ corresponding to the Lasso and ridge case, respectively.
Better solutions can be obtained for $q<1$, however the resulting optimization
problem is no longer convex and therefore more difficult to solve.

An alternative Bayesian approach to obtain a sparse solution using an $L_0$
penalty was proposed by \cite{george1993}. They introduce $n$ variational
selector variables $s_i$ such that the prior distribution over $w_i$ is a
mixture of a narrow (spike) and wide (slab) Gaussian distribution, both
centered on zero. The posterior distribution over $s_i$ indicates whether the
input feature $i$ is included in the model or not.  Since the number of subsets
of features is exponential in $n$, for large $n$ one cannot compute the
solution exactly. In addition, the posterior is a complex high dimensional
distribution of the $w_i$ and the other (hyper) parameters of the model.  The
computation of the posterior requires thus the use of MCMC sampling
\cite{george1993} or  a variational Bayesian approximation
\cite{titsias2012,logsdon2010,Yoshida,lobato}.


Although Bayesian approaches tend to over fit less than a maximum likelihood or
maximum a posteriori method (MAP approach), they also tend  to be relatively
slow.  Here we propose a partial Bayesian approach, where we apply a
variational approximation to integrate out the binary (selector) variables in combination
with a MAP approach for the remaining parameters.  For clarity, we analyse this
idea in its most simple form, in the absence of (hierarchical) priors. Instead,
we infer the sparsity prior through cross validation. As we will motivate
below, we call the method the Variational Garrote (VG).

The paper is organized as follows. In section~\ref{variational} we
introduce the model and we derive the variational approximation.
We show that the combination of the variational approximation and
$L_0$ regularization has the effect of making the problem effectively
of maximal rank by introducing a 'variational ridge term'. As a
result, the solution is well defined even when $p<n$ as long as the
number of predictive features is less than $p$ (which is controlled
by the sparsity prior).

To gain further insight,  in section~\ref{univariate} we
study the case when the design matrix is orthogonal. In this case the
solution can be computed exactly in closed form with no need to resort to approximations. 
In the variational approximation, we show for the uni-variate case that the solution is
either unique or has two solutions, depending on the input-output
correlations, the number of samples $p$ and on the sparsity prior
$\gamma$. We derive a phase plot and show that the solution is
unique, when the sparsity prior is not too strong {\em or} when the
input-output correlation is not too large.  The input-output behavior
of the VG is shown to be close to optimal as a smoothed version of
hard feature selection. We argue that this behavior also holds
in the multi-variate case.

In section~\ref{numerical} we compare the VG with a number of other MAP
methods, such as Lasso and ridge regression and with the paired mean field
method (PMF) \cite{titsias2012}, a recently proposed variational bayesian
method.  We show that the VG and PMF significantly outperform the Lasso and
ridge regression on a large number of different examples both in terms of the
accuracy of the solution, as well as in prediction error.  In addition, we show
that the VG do not suffer from the inconsistency of the Lasso method when the
input correlations are large.  We show in detail how all methods compare as a
function of the level of noise, the sparsity of the target solution, the number
of samples and the number of irrelevant predictors.   Globally, VG is
significantly faster than PMF and tends to perform better as the
problems become denser and in problems with strongly correlated inputs.


\section{The variational approximation}
\label{variational}
Consider the regression model of the form 
\footnote{We assume from here on without loss of generality that 
$\frac{1}{p}\sum_{\mu=1}^p x_i^\mu=\frac{1}{p}\sum_{\mu=1}^p y^\mu=0$}
\bea
y^\mu=\sum_{i=1}^n w_i s_i x^\mu_i+\xi^\mu\qquad \sum_{i=1}^n s_i \le
t
\label{vg}
\eea
with $s_i=0,1$.  The bits $s_i=1$ will identify the predictive inputs
$i$.  
Using a Bayesian description, and denoting the data by $D: \{\vx^\mu,y^\mu\},
\mu=1,\ldots,p$, the likelihood term is given by 
\begin{align}
\label{likelihood}
p(y|\vx,\vs,\vw,\beta)&=\sqrt{\frac{\beta}{2\pi}}\exp\left(-\frac{\beta}{2}\left(y-\sum_{i=1}^n
w_i s_i x_i\right)^2\right)\nonumber\\
p(D|\vs,\vw,\beta)&=\prod_\mu p(y^\mu|\vx^\mu,\vs,\vw,\beta)\notag\\
&=\left(\frac{\beta}{2\pi}\right)^{p/2}\exp\left(-\frac{\beta p}{2}\left(\sum_{i,j=1}^n
s_i s_j w_i w_j \chi_{ij}-2\sum_{i=1}^n w_i s_i b_i +\sigma_y^2\right)
\right)
\end{align}
with 
$b_i=\frac{1}{p}\sum_\mu x_i^\mu y^\mu,
\sigma_y^2=\frac{1}{p}\sum_\mu (y^\mu)^2,
\chi_{ij}=\frac{1}{p}\sum_\mu x_i^\mu x_j^\mu$.

We should also specify prior distributions over $\vs,\vw,\beta$.
For concreteness, we assume that the prior over $\vs$ is factorized
over the individual $s_i$, each with identical prior probability:
\begin{align}
\label{prior}
p(\vs|\gamma)& =\prod_{i=1}^n p(s_i|\gamma)  &
p(s_i|\gamma)& =\frac{\exp\left(\gamma s_i\right)}{1+\exp(\gamma)}
\end{align}
with $\gamma$ given which specifies the sparsity of the solution. 
We denote by $p(\vw,\beta)$ the prior over 
the inverse noise variance $\beta$ and the feature weights $\vw$. We will 
leave this prior unspecified since its choice does not affect the variational approximation. 
\footnote{
It can be shown that the regression model specified by Eqs.~\ref{likelihood} and~\ref{prior} is identical to the spike and slab model, with the difference that the latter usually contains a (Gaussian) prior over the $w_i$ which could also be added in the above representation\cite{titsias2012}. See appendix~\ref{sec:pmf} for details.

%
}

The posterior becomes
\bea
p(\vs,\vw,\beta|D,\gamma)=\frac{p(\vw,\beta)p(\vs|\gamma)p(D|\vs,\vw,\beta)}{p(D|\gamma)}
\label{posterior}
\eea
Computing the MAP estimate or computing statistics from the
posterior is complex in particular due to the discrete nature or $\vs$.  We propose
to compute a variational approximation to the marginal posterior
$p(\vw,\beta|D,\gamma)=\sum_{\vs} p(\vs,\vw,\beta|D,\gamma)$ and computing the
MAP solution with respect to $\vw,\beta$. Since $p(D|\gamma)$ does not depend on
$\vw,\beta$ we can ignore it. 

The posterior distribution Eq.~\ref{posterior} for given $\vw,\beta$ is a
typical Boltzmann distribution involving terms linear and quadratic in $s_i$.
It is well-known that when the effective couplings $w_i w_j \chi_{ij}$ are
small, one can obtain good approximations using methods that originated in the
statistical physics community and where $s_i$ denote binary spins. Most
prominently, one can use the mean field or variational approximation
\cite{jordan1998}, the TAP approximation \cite{kap99e} or belief propagation
(BP) \cite{murphy99}.  For introductions into these methods also see
\cite{opper2001a,wainwright2008}.  Here, we will develop a solution based on
the simplest possible variational approximation and leave the possible
improvements using BP or structured mean field approximations to the future. 

We approximate the sum by the variational bound using Jensen's inequality.
\begin{align}
\label{bound}
\log \sum_{\vs}p(\vs|\gamma)p(D|\vs,\vw,\beta)& \ge -\sum_{\vs} q(\vs) \log
\frac{q(\vs)}{p(\vs|\gamma)p(D|\vs,\vw,\beta)}\notag\\&=-F(q,\vw,\beta)
\end{align}
$q(\vs)$ is called the variational approximation and can be any
positive probability distribution on $\vs$ and $F(q,\vw,\beta)$ is called the variational
free energy.
The optimal $q(\vs)$ is found by
minimizing $F(q,\vw,\beta)$ with respect to $q(\vs)$ so that the
tightest bound - best approximation - is obtained.

In order to be able to compute the variational free energy efficiently,
$q(\vs)$ must be a tractable probability distribution, such as a chain or a
tree with limited tree-width \cite{barber98b}. Here we consider the simplest
case where $q(\vs)$ is a fully factorized distribution: $q(\vs)=\prod_{i=1}^n q_i(s_i)$ with $q_i(s_i)=m_i s_i
+(1-m_i)(1-s_i)$, so that $q$ is fully specified by the expected values
$m_i=q_i(s_i=1)$, which we collectively denote by $\vm$.
The expectation values with respect to $q$ can now be easily evaluated
and the result is
\begin{align}
\label{F}
F&=
\frac{\beta p}{2}\left(\sum_{i,
j}^n m_i m_j w_i w_j \chi_{ij}+\sum_i m_i(1-m_i) w_i^2
\chi_{ii}-2\sum_{i=1}^n m_i w_i b_i
+\sigma_y^2\right)\nonumber\\
&-\gamma\sum_{i=1}^n m_i +\sum_{i=1}^n \left(m_i \log m_i+(1-m_i)\log
(1-m_i)\right)-\frac{p}{2}\log\frac{\beta}{2\pi}
\end{align}
where we have omitted terms independent of $m,\beta,w$. 
The first line is due to the 
likelihood term, the
second line is due to the prior on $\vs$ and the entropy of
$q(\vs)$.
The approximate marginal posterior is then
\begin{align*}
p(\vw,\beta|D,\gamma)&\propto p(\vw,\beta)\sum_{\vs}p(\vs|\gamma)p(D|\vs,\vw,\beta)\\
& \approx p(\vw,\beta)
\exp(-F(\vm,\vw,\beta,\gamma))
\end{align*}

We can compute the variational approximation $\vm$ for given
$\vw,\beta,\gamma$
by minimizing $F$ with respect to $\vm$. 
In addition, 
$p(\vw,\beta|D,\gamma)$ needs to be maximized with respect to
$\vw,\beta$. 
Note, that the variational approximation only depends on the likelihood term and
the prior on $\gamma$, since these are the only terms that depend on $\vs$.
Thus, for given $\vw$, the variational approximation does not depend on the
particular choices for the prior $p(\vw,\beta)$.
For concreteness, we assume a flat prior $p(\vw,\beta)\propto 1$. We set
the derivatives of $F$ with
respect $\vm,\vw,\beta$ equal to zero. This gives the
following set of fixed point equations:
\bea
m_i&=&\sigma\left(\gamma+\frac{\beta p}{2}w_i^2 \chi_{ii}\right)
\label{m}\\
\vw&=&(\chi')^{-1} \vb \label{v}\qquad
\chi'_{ij}=\chi_{ij}m_j+(1-m_j)\chi_{jj}\delta_{ij}\\
\frac{1}{\beta}&=&\sigma_y^2-\sum_{i=1}^n m_i w_i b_i
\label{beta}
\eea
with $\sigma(x)=(1+\exp(-x))^{-1}$ and where in Eq.~\ref{beta} we have
used Eq.~\ref{v}.
Eqs.~\ref{m}-\ref{beta} provide the final solution. They can be solved by fixed point iteration as outlined in
Algorithm~\ref{algorithm}: Initialize $\vm$ at random. Compute $\vw$ by
solving the linear system Eq.~\ref{v} and $\beta$ from Eq.~\ref{beta}. 
Compute a new solution for $\vm$ from Eq.\ref{m}.

Within the variational/MAP approximation the predictive model is given by 
\bea
y&=&\sum_i m_i w_i x_i +\xi\label{prediction}
\eea
with $\av{\xi^2}=1/\beta$ and $\vm,\vw,\beta$ as estimated by the above procedure. 
Eq.~\ref{prediction} has some similarity with Breiman's non-negative
Garrote method \cite{breiman93}. It computes the solution in a two step approach: it computes first $w_i$ using OLS and then finds $m_i$
by minimizing
\beaa
\sum_\mu \left(y^\mu -\sum_{i=1}^n x_i^\mu w_i m_i\right)^2\quad
\mathrm{subject~to}\quad  m_i\ge 0\quad \sum_i m_i \le t
\eeaa
Because of this similarity, we refer to our method as the variational
Garrote (VG).
Note, that because of the OLS step the non-negative garrote requires that $p\ge
n$.  Instead, the variational solution Eqs.~\ref{m}-\ref{beta} computes the
entire solution in one step (and as we will see does not require $p\ge n$).

Let us pause to make some observations about this solution. One
might naively expect that the variational approximation would simply
consist of replacing $w_i s_i$ in Eq.~\ref{vg} by its variational
expectation $w_i m_i$. If this were the case, $\vm$ would
disappear entirely from the equations and one would expect in Eq.~\ref{v} the OLS solution with the
normal input covariance matrix $\chi$ instead of the new matrix
$\chi'$ (note, that in the special case that $m_i=1$ for all $i$,
$\chi'=\chi$ and Eq.~\ref{v} does reduce to the OLS
solution).
Instead, $\vm$ and
$\vw$ are both to be optimized, giving in general a
different solution than the OLS solution \footnote{The technical reason that
this does not occur is that in the computation of the expectation with respect to the distribution $q$ 
that results in Eq.~\ref{F}
one has $\av{s_i s_j}=m_i m_j$ for $i\ne j$, but
$\av{s_i^2}=\av{s_i}=m_i$.}.

When $m_i<1$, $\chi'$ differs from $\chi$ by rescaling with $m_i$ and adding a positive
diagonal to it, a 'variational ridge'. This is similar to the mechanism of ridge regression,
but with the important difference that the diagonal term depends
on $i$ and is dynamically adjusted depending on the solution for
$\vm$.  Thus, the sparsity prior together with variational approximation
provides a mechanism that solves the rank problem.
When all $m_i<1$, $\chi'$ is of maximal rank. Each $m_i$ that approaches 1,
reduces the rank by one.
Thus, if $\chi$ has rank $p<n$, $\chi'$ can be still of rank $n$
when no more than $p$ of the $m_i=1$, the remaining $n-p$ of the $m_i<1$
making up for the rank deficiency. Note, that the size of $m_i$ (and thus the rank
of $\chi'$) is
controlled by $\gamma$ through Eq.~\ref{m}.

In the above procedure, we compute the VG solution for fixed $\gamma$
and choose its optimal value through cross validation on independent
data \cite{mitchell1988}. This has the advantage that our result
is independent of our (possibly incorrect) prior belief.

But another important advantage of varying $\gamma$ manually is
that it helps to avoid local minima.  When we increase $\gamma$
from a negative value $\gamma_\mathrm{min}$ to a maximal value  $\gamma_\mathrm{max}$ 
in small steps, we obtain a sequence
of solutions with decreasing sparseness. These solutions will better
fit the data and as a result $\beta$ increases with $\gamma$. Thus,
increasing $\gamma$ implements an annealing mechanism where we
sequentially obtain solutions at lower noise levels. We found
empirically that this approach is effective to reduce the problem
of local minima. To further deal with the effect of hysteresis (see
section~\ref{univariate}) we
recompute the solution  from $\gamma_\mathrm{max}$ down to
$\gamma_\mathrm{min}$ and choose the solution with lowest free
energy.

The minimal value of $\gamma$ is chosen as the largest value such that
$m_i=\epsilon$, with $\epsilon$ small. We find from
Eqs.~\ref{m}-\ref{beta} that 
\bea
\gamma_\mathrm{min} =-\frac{p
b_i^2\chi_{ii}}{2\sigma_y^2}+\sigma^{-1}(\epsilon)+\cO(\epsilon)\label{gamma_min}
\eea
with $\sigma^{-1}(x)=\log(x/(1-x))$.
We heuristically set the maximal value of $\gamma$ as well as the step
size. 

In appendix~\ref{dual} we provide an alternative fixed point iteration
scheme that is more efficient in the large $n$ small $p$ limit.
Whereas Eqs.~\ref{m}-\ref{beta} require the repeated solution of
a $n$-dimensional linear system, the dual formulation,
Eqs.~\eqref{m},\eqref{w2},\eqref{A}-\eqref{lambda1}, requires the repeated
solution of a $p$ dimensional linear system.
Algorithm \ref{algorithm} summarizes the VG method.

\begin{algorithm}
\SetKwFunction{VG}{VG}\SetKwFunction{VG}{VG}
 \SetKwInOut{Input}{input}\SetKwInOut{Output}{output}
 \SetKwInOut{Function}{function}
 \SetKwInOut{Output}{output}
\Input{Data $D: \{\vx^\mu,y^\mu\},\mu=1,\ldots,p$ ; $\epsilon$ and step-size $\Delta \gamma$}
\Output{$\vw,\vm,\beta,\gamma$ solution with minimal cross validation error }
\nl Preprocess data such that $\sum_\mu x_i^\mu=\sum_\mu y^\mu=0$ and partition $D$ in $D^{\text{train}}$, $D^{\text{val}}$\\
\nl Compute $b_i =\frac{1}{p}\sum_\mu x_i^\mu y^\mu$ and if $n<p$ compute $\chi_{ij}=\frac{1}{p}\sum_\mu x_i^\mu x_j^\mu$\\
\nl Compute $\gamma_\mathrm{min}$ from $\epsilon$ 
 and $\gamma_\mathrm{max}$ from $\gamma_\mathrm{min}$ and $\Delta\gamma$ \\
\nl \For(// FORWARD PASS){$\gamma=\gamma_\mathrm{min}:\Delta \gamma:\gamma_\mathrm{max}$}
{
\nl $\eta\leftarrow 1$\\
\nl \While{not converged}{
\nl Compute $\vw,\beta$ from Eqs.~\eqref{v}-\eqref{beta} ($n<p$)
	or Eqs.~\eqref{w2},~\eqref{A}-\eqref{lambda1} ($n>p$)\;
\nl Compute $\vm'$ using a smoothed version of Eq.~\eqref{m}: $m_i'\leftarrow (1-\eta) m_i + \eta \sigma(\ldots)$\\
\nl \If{$\max_i |m_i'-m_i|>0.1$}{
\nl $\eta\leftarrow \eta/2$}
\nl $\vm\leftarrow\vm'$\\
}
\nl    Store solution $(\vw_1,\vm_1,\beta_1)_\gamma$ and $F_1(\gamma)\leftarrow F((\vw_1,\vm_1,\beta_1)_\gamma)$ from Eq.~\eqref{F}\\
}
\nl \For(// BACKWARD PASS){$\gamma=\gamma_\mathrm{max}:-\Delta \gamma:\gamma_\mathrm{min}$}
{
\nl As $\mathbf{5}-\mathbf{11}$\\
\nl    Store solution $(\vw_2,\vm_2,\beta_2)_\gamma$ and $F_2(\gamma)\leftarrow F((\vw_2,\vm_2,\beta_2)_\gamma)$ from Eq.~\eqref{F}\\
}
\nl \For{$\gamma=\gamma_\mathrm{min}:\Delta \gamma:\gamma_\mathrm{max}$}{
\nl Choose solution $(\vw,\vm,\beta)_\gamma$ that has minimal $F_{1,2}(\gamma)$\\ 
\nl     Compute cross validation error on $D^{\text{val}}$ using Eq.~\eqref{prediction} \\
 }
\nl Select $\vw,\vm,\beta,\gamma$ with minimal cross validation error\\
\caption{The Variational Garrote algorithm.}\label{algorithm}
\end{algorithm}

\section{Orthogonal and uni-variate case}
\label{univariate}
In order to obtain further insight in the solution, consider the case in
which the inputs are uncorrelated: $\chi_{ij}=\delta_{ij}$. In this
case, we can derive the MAP solution of Eq.~\ref{posterior} exactly, without the need to
resort to the variational approximation. Eq.~\ref{posterior}
reduces to a distribution that factorizes over $i$ with
log probability proportional to
\beaa
L=\frac{p}{2}\log\beta-\frac{\beta p}{2}\left(\sum_{i=1}^n s_i
(w_i^2-2w_i b_i)
 +\sigma_y^2\right)+\gamma \sum_{i=1}^n s_i
\eeaa
Maximizing wrt $w_i,\beta$ yields $w_i=b_i$,
$\beta^{-1}=\sigma_y^2-\sum_{i=1}^n s_i 
b_i^2$ and
\beaa
L=\frac{p}{2}\log\beta+\sum_{i=1}^n s_i \left(\frac{\beta
p}{2}b_i^2+\gamma\right)-\frac{\beta p}{2} \sigma_y^2
\eeaa
Assume without loss of generality that $b_i^2$ are sorted in decreasing order. $L$ is maximized by setting $s_i=1$ when $\frac{\beta 
p}{2}b_i^2+\gamma>0$ and $s_i=0$ otherwise. Thus, the optimal solution
is $s_{1:k}=1, s_{k+1:n}=0$, $\beta^{-1}=\sigma_y^2-\sum_{i=1}^k  
b_i^2$ with $k$ the smallest integer such that 
\bea
\frac{\beta p}{2}b_{k+1}^2+\gamma <0
\label{switch_exact}
\eea
By varying $\gamma$ from small to large, we find a sequence of
solutions with decreasing sparsity.

In the variational approximation the solution is very similar but not identical. Eq.~\ref{v} gives the same solution
$w_i=b_i$. Eqs.~\ref{m} and~\ref{beta} become 
\beaa
m_i&=&\sigma\left(\gamma+\frac{\beta p}{2}b_i^2\right)\\
1/\beta&=&\sigma_y^2-\sum_i b_i^2 m_i
\eeaa
which we can interpret as the variational approximations of
Eq.~\ref{switch_exact}, with $m_{1:k}\approx 1$ and $m_{k+1:n}\approx
0$.
The term $\sum_i b_i^2 m_i$ is the explained variance and is subtracted from
the total output variance to give an estimate of the noise variance
$1/\beta$. 

Note that the posterior is factorized in $s_i$, the variational approximation is not identical to the exact map solution Eq.~\ref{switch_exact}, although the results are very similar. The relation is $s_i=0 \Leftrightarrow m_i < 0.5$ and $s_i=1 \Leftrightarrow m_i > 0.5$.

In order to further analyze the variational solution, we consider the 
$1$-dimensional case. The variational equations become
\bea
m&=& \sigma\left(\gamma + \frac{p}{2}
\frac{\rho}{1-\rho m}\right)=f(m)\label{univ1}\\
\frac{1}{\beta}&=&\sigma^2_y(1-m \rho)\label{univ2}
\eea
with $\rho=b^2/\sigma_y^2$ the squared correlation coefficient. 

In Eq.~\ref{univ1}, we have eliminated $\beta$ and we must find a
solution for $m$ for this non-linear equation. We see that it depends
on the input-output correlation $\rho$, the number of samples $p$ and
the sparsity $\gamma$. 
For $p=100$, the solution for different $\rho,\gamma$ is
illustrated in figure~\ref{file4b} (see appendix~\ref{appendix:univ}).
\begin{figure}
\bc
\includegraphics[width=0.6\textwidth]{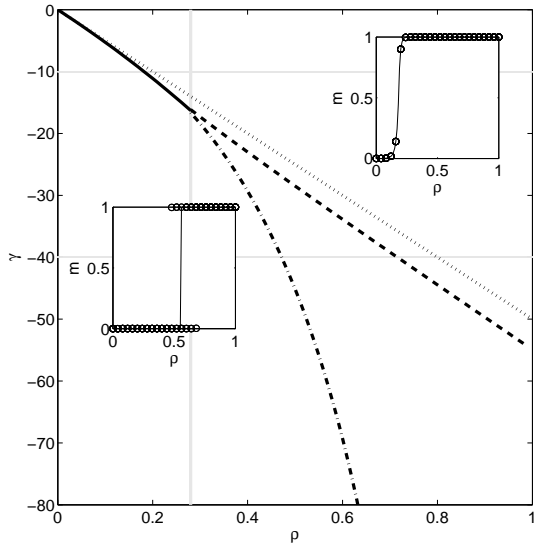}
\ec
\caption{
Phase plot $\rho,\gamma$ for $p=100$ giving the different solutions for $m$.
Dashed and dot-dashed lines for $\rho>\rho^*=0.28$ are from
Eq.~\ref{appendix_univ2} where two solutions for $m$ exist. Solid line for
$\rho<\rho^*$ is the solution for
$\gamma$ when $m=1/2$, to indicate the transition from the unique solution 
$m\approx 0$ to the unique solution $m\approx 1$.
Dotted line is the exact transition from $s=0$ to $s=1$ from
Eq.~\ref{switch_exact}.
Insets indicate solutions for $m$ versus $\rho$ for $\gamma=-10,p=100$ (top-right)
and for $\gamma=-40,p=100$ (bottom-left).
In the lower left corner of the insets, the unique solution $m\approx 0$ is found. In the top
right corner, the unique solution $m\approx 1$ is found. Between the dot-dashed and the dashed line, the two variational solutions $m\approx 0$ and $m\approx 1$ co-exist.
}
\label{file4c}
\end{figure}
Eq.~\ref{univ1} has one or three solutions for $m$, depending on the values of
$\gamma,\rho,p$. 
The three solutions correspond to two local
minima and one local maximum of the free energy $F$.
For $\gamma=-40$ and $\gamma=-10$, we plot the stable solution(s) for different values of
$\rho$ in the inserts in fig.~\ref{file4c}. 
The best variational solution for $m$ is given by the
solution with the lowest free energy, indicated by the solid lines in
the inserts in fig.~\ref{file4c}. 

Fig.~\ref{file4c} further shows the phase plot of $\gamma,\rho$ that indicates that the
variational solution is unique for $\gamma>\gamma^*$ or for $\rho<\rho^*$.
The solid line for $0<\rho<\rho^*$ 
in fig.~\ref{file4c}
indicates a smooth (second
order) phase transition from $m=0$ to $m=1$.
For $\rho>\rho^*$, the
transition from $m=0$ to $m=1$ is discontinuous: for each $\rho$ there is 
a range of values of
$\gamma$ where two variational solutions $m\approx 0$ and $m\approx 1$ co-exist.
For comparison, we also show the line $\gamma=-p\rho/2$ that separates the
solution $s=0$ and $s=1$ according the the exact (non-variational) solution
Eq.~\ref{switch_exact}.

The multi-valued variational solution results in a hysteresis effect.  When the
solution is computed for increasing $\gamma$, the $m\approx 0$ solution is
obtained until it no longer exists. If the sequence of solutions is computed
for decreasing $\gamma$ the $m\approx 1$ solution is obtained for values of
$\gamma$ where previously the $m\approx 0$ solution was obtained. 

From this simple one-dimensional case we may infer
that the variational
approximation is relatively easy to compute in the uni-modal region (small $\rho$
or $\gamma$ not too negative) and becomes more inaccurate in the region where
multiple minima exist (region between the dot-dashed and dashed lines in fig.~\ref{file4c}) . 

It is interesting to compare the uni-variate solution of the variational garrote with ridge
regression, Lasso or Breiman's Garrote, which was previously done for the
latter three methods in \cite{tibshirani96}. Suppose that data are generated from
the model $y=w x + \xi$ with $\av{\xi^2}=\av{x^2}=1$. We compare the solutions as a
function of $w$. The OLS solution is approximately given by $w_\mathrm{ols}\approx
\av{x y}=w$, where we ignore the statistical deviations of order $1/p$ due to the finite
data set size. Similarly, the ridge
regression solution is given by $w_\mathrm{ridge}\approx \lambda w$, with $0<
\lambda < 1$ depending on the ridge prior.
The Lasso solution (for non-negative $w$) is given by
$w_\mathrm{lasso}=(w-\gamma)^+$ \cite{tibshirani96}, with $\gamma$ depending on the
$L_1$ constraint.
Breiman's Garrote solution is given by $w_\mathrm{garrote}=(1-\frac{\gamma}{w^2})^+ w$
\cite{tibshirani96}, with $\gamma$ depending on the 
$L_1$ constraint. The VG solution is given by $w_\mathrm{vg}=m w$,
with $m$ the solution of Eq.~\ref{univ1}. Note, that the VG solution
depends, in addition to $w,\gamma$, on the unexplained variance $\sigma_y^2$ and the number of
samples $p$, whereas the other methods do not.

\begin{figure}
\bc
\includegraphics[width=0.5\textwidth]{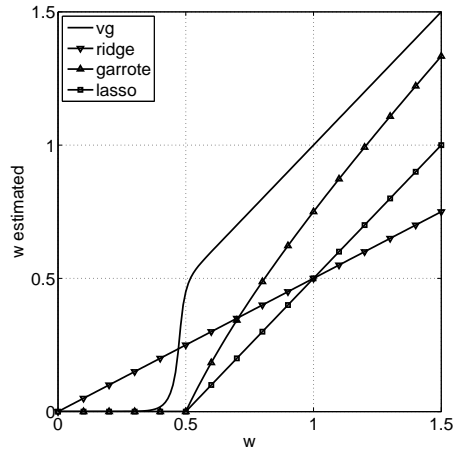}
\ec
\caption{Uni-variate solution for different regression methods. All methods yield a
shrinked solution (deviation from diagonal line). Variational Garrote (VG) with
$\gamma=-10, p=100$ and $\sigma_y^2=1$. Ridge regression with $\lambda=0.5$. Garrote with
$\gamma=1/4$. Lasso with $\gamma=1/2$.
}\label{file5a}
\end{figure}

The qualitative difference of the solutions is shown in fig.~\ref{file5a}.
The ridge regression solution is off by a constant multiplicative factor. 
The Lasso solution is zero for small $w$ and for larger $w$ gives a solution that 
is shifted downwards by a constant factor.
Breiman's Garrote is identical to the Lasso for small $w$ and shrinks less for larger
$w$. The VG gives an almost ideal behavior and can
be interpreted as a soft version of variable selection: For small
$w$ the solution is close to zero and the variable is ignored, and
above a threshold it is identical to the OLS solution. 

The qualitative nature of the phase plot fig.~\ref{file4c} and the
input-output behavior fig.~\ref{file5a} extends to the multi-variate
orthogonal case. The symmetry breaking of feature $i$ is independent
of all other features, except for the term $\delta= \sum_{j\ne
i}b_j^2 m_j$ that enters through $\beta$. If we increase $\gamma$, $\delta$ increases in steps
each time that one of the features $j$ switches from $m_j\approx 0$ to
$m_j\approx 1$.
Thus $\delta$ is constant almost always, except at the step points.
Since the critical values of $\rho$ and $\gamma$ depend in a simple
way on $\delta$, the phase plot for the multivariate orthogonal
case is qualitatively the same as for the uni-variate case.

\section{Numerical examples}
\label{numerical}
In the following examples, we compare the VG with Lasso, ridge regression and
in some cases, with the paired mean field approach (PMF) \cite{titsias2012}.

For most of the examples, we generate a training set, a validation set and a
test set.  Inputs are generated from a zero mean multi-variate Gaussian
distribution with specified covariance structure.  We generate outputs
$y^\mu=\sum_i \hat{w}_i x_i^\mu +d\xi^\mu$ with $d\xi^\mu \in
\cN(0,\hat{\sigma})$ and $\hat{w}_i$ depending on the problem.

For VG, ridge regression and Lasso, we optimize the model parameters on the
training set and, when necessary, optimize the hyper parameters ($\gamma$ in
the case of VG, $\lambda$ in the case of ridge regression and Lasso) that
minimize the quadratic error on the validation set.  For the Lasso, we used the
method described in \cite{friedman2009}
\footnote{
\url{http://www-stat.stanford.edu/~tibs/glmnet-matlab/}.}. 

Comparison with PMF is performed using the software available online for the
regression case with one-dimensional output
\footnote{
\url{http://www.well.ox.ac.uk/~mtitsias/software.html}.}.
Since PMF optimizes hyperparameters as well, we merge both training and
validation sets and the resulting dataset is used as input for the PMF method.
This ensures that all methods use the same data for parameter estimation.

We define the solution vector for a given method as $\vv$.  For VG, the
components are $v_i\equiv m_i w_i$.  In the case of PMF, $m_i$ corresponds to
the spike-and-slab variational posterior and $w_i$ to the variational mean for
the weights \footnote{ The notation in~\cite{titsias2012} uses $\tilde w_i$ for
$w_i$ and $\gamma_i$ for $m_i$.  }.  For Ridge and Lasso $v_i\equiv w_i$.



\subsection{Small Example 1}
In the first example, we take independent inputs $x_i^\mu \in
\cN(0,1)$ and a teacher weight vector with only one non-zero entry:
$\hat{w}=(1,0,\ldots,0)$, $n=100$ and $\hat{\sigma}=1$. The training set size
$p=50$, validation set size $p_v=50$ and test set size $p_t=400$. 
\begin{figure}
\bc
\includegraphics[width=0.65\textwidth]{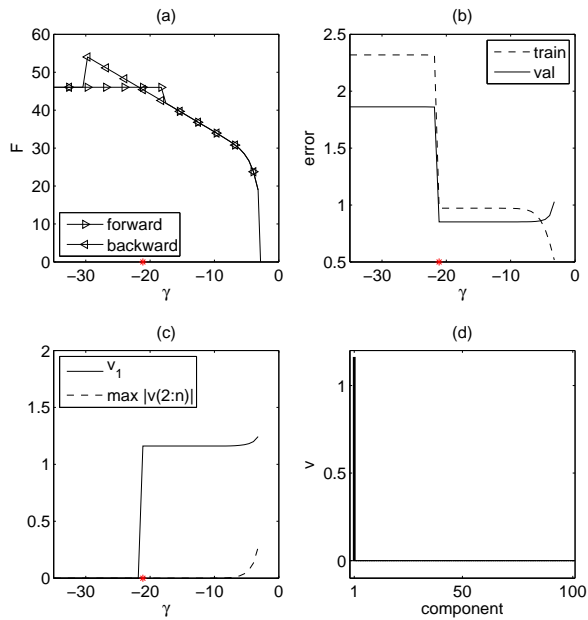}
\ec
\caption{Top left (a): Minimal variational free energy versus $\gamma$. The two
curves correspond to warm start solution from small to large $\gamma$ ('forward') and from
large to small $\gamma$ ('backward') (see also Algorithm~\ref{algorithm}).
Top right (b): Training and validation error versus $\gamma$. The optimal $\gamma$
minimizes the validation error.
Bottom left (c): Solution $v_1=m_1 w_1$ and $\max_{i=2:n} |m_i w_i|$. The correct
solution is found in the range $\gamma\approx-20$ to $\gamma\approx -5$.
Bottom right (d): Optimal solution $v_i=w_i m_i$ versus $i$.
}
\label{file7a}
\end{figure}
We choose $\epsilon=0.001$ in Eq.~\ref{gamma_min},
$\gamma_\mathrm{max}=0.02
\gamma_\mathrm{min}, \Delta \gamma=-0.02 \gamma_\mathrm{min}$
(see Algorithm~\ref{algorithm} for details).

Results for a single run of the VG are shown in fig.~\ref{file7a}.  In
fig.~\ref{file7a}a, we plot the minimal variational free energy $F$ versus
$\gamma$ for both the forward and backward run.  Note, the hysteresis effect
due to the local minima. For each $\gamma$, we use the solution with the lowest
$F$.  In fig.~\ref{file7a}b, we plot the training error and validation error
versus $\gamma$. The optimal $\gamma\approx -21$ is denoted by a star and the
corresponding $\sigma=1/\sqrt{\beta}=1.05$.  In fig.~\ref{file7a}c, we plot the
non-zero component $v_1=m_1 w_1$ and the maximum absolute value of the
remaining components versus $\gamma$. Note the robustness of the VG solution in
the sense of the large range of $\gamma$ values for which the correct solution
is found.  In fig.~\ref{file7a}d, we plot the optimal solution $v_i=m_i w_i$
versus $i$. 

\begin{figure}
\bc
\includegraphics[width=0.8\textwidth]{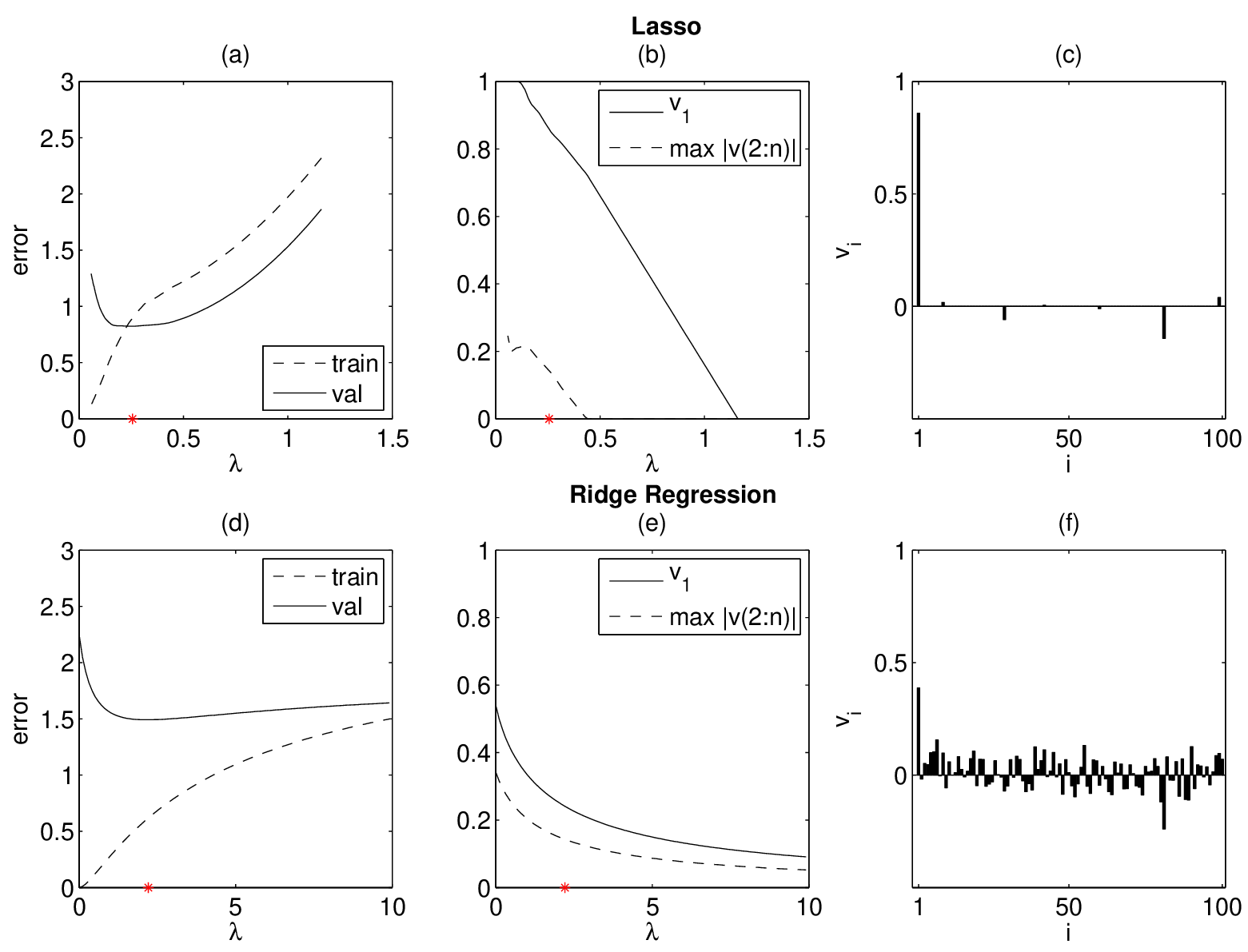}
\ec
\caption{Regression solution for Lasso and ridge regression
for same data set as in fig.~\ref{file7a}. 
Top row (a,b,c): Lasso. Bottom row (d,e,f): Ridge regression.
Left column (a,d): training and validation errors versus $\lambda$.
Middle column (b,e): Solution for the non-zero feature $v_1$ and the
zero-features $\max_{i=2:n} |v_i|$. 
Right column (c,f): Optimal Lasso and ridge regression solution $v_i$ versus $i$.
}
\label{file7b}
\end{figure}

In fig.~\ref{file7b} we show the Lasso (top row) and ridge regression (bottom row) 
results for the same
data set. The optimal value for $\lambda$ minimizes the validation error (star). 
In fig.~\ref{file7b}b,c we see that the Lasso selects a number of 
incorrect features as well. Fig. ~\ref{file7b}b also shows that the
Lasso solution with a larger $\lambda$ in
the range $0.45<\lambda< 0.95$ could select the single correct feature, but
would then estimate $\hat{w}_1$ too small due to the large shrinkage effect. 
Ridge regression gives very bad results. The non-zero feature is too
small and the remaining features have large values. Note from fig.~\ref{file7b}e,
that ridge regression yields a non-sparse solution for all values of $\lambda$.

\begin{table}
\bc
\begin{tabular}{l|r|r|r|r|r}
			&Train					&Val				&Test				&\# non-zero		&$\|\delta \vv\|_1$ 	\\\hline
Ridge		&$  0.60 \pm   0.43$	&$  1.72 \pm   0.39$&$  1.80 \pm   0.12$&$-$					&$  3.97 \pm   1.23$\\
Lasso		&$  0.78 \pm   0.26$	&$  1.07 \pm   0.20$&$  1.17 \pm   0.20$&$  8.65 \pm   6.75  $&$  0.80 \pm   0.57$\\
PMF       &$  -$    &$-$                  &$  1.02 \pm   0.10$&$   1.5 \pm   1.19 $&$  0.33 \pm   0.37$\\
VG			&$  0.85 \pm   0.22$	&$  0.96 \pm   0.17$&$  1.01 \pm   0.10$&$ 1.20 \pm   0.52$&$  0.31 \pm   0.30$ \\
True		&$  0.93 \pm   0.14$	&$  0.87 \pm   0.20$&$  0.98 \pm   0.04$&$1$			&$0$					
\end{tabular}
\ec
\caption{Results for Example 1 averaged over 20 instances.
Train is mean squared error (MSE) on the training set.  Val is MSE on the
validation set.  Test is MSE on the test set.  \#~non-zero is the number of
non-zero elements in the Lasso solution and $\sum_{i=1}^n (m_i>0.5)$ for
VG and PMF.  $\|\delta\vv\|_1=\sum_{i=1}^n |v_i-\hat{w}_i|$.
}
\label{table1}
\end{table}
Table~\ref{table1} shows that the VG significantly outperforms the Lasso method
and ridge regression both in terms of prediction error, the accuracy of the
estimation of the parameters and the number of non-zero parameters. In this
simple example, there is no significant difference in the prediction error of
Lasso, PMF and VG, but the Lasso solution is significantly less sparse.  There
is no significant difference between the solutions found by PMF and VG.

\subsection{Small Example 2}
In the second example, we consider the effect of correlations in
the input distribution. Following \cite{tibshirani96} we generate input data
from a multi-variate Gaussian distribution with covariance matrix
$\chi_{ij}=\zeta^{|i-j|}$, with $\zeta=0.5$. In addition, we choose multiple
features non-zero: $\hat{w}_i=1, i=1,2,5,10,50$ and all other $\hat{w}_i=0$. We
use $n=100, \hat{\sigma}=1$ and $p/p_v/p_t=50/50/400$.  In table~\ref{table2}
we compare the performance of the VG, Lasso, ridge regression and PMF
on 20 random instances.  
\begin{table}
\bc
\begin{tabular}{l|r|r|r|r|r}
			&Train					&Val				&Test				&\# non-zero		&$\|\delta\vv\|_1$ 	\\\hline
Ridge		&$  0.32 \pm   0.27$	&$  3.30 \pm   0.67$&$  3.46 \pm   0.31$&$-$					&$ 11.09 \pm   0.93$\\
Lasso		&$  0.75 \pm   0.37$	&$  1.39 \pm   0.37$&$  1.48 \pm   0.29$&$ 16.30 \pm   6.60$&$  2.08 \pm   0.87$\\
PMF       &$  -$    &$-$                  &$  1.06 \pm   0.11$&$   5.15 \pm   0.49 $&$  0.67 \pm   0.35$\\
VG			&$  0.80 \pm   0.25$	&$  1.13 \pm   0.31$&$  1.15 \pm   0.21$&$  5.05 \pm   0.51$&$  0.83 \pm   0.54$\\
True		&$  0.93 \pm   0.14$	&$  0.87 \pm   0.20$&$  0.98 \pm   0.04$&5&0
\end{tabular}
\ec
\caption{Results for Example 2. For definitions see caption of Table~\ref{table1} above.
}
\label{table2}
\end{table}
We see that the VG and PMF significantly outperform the Lasso method and ridge
regression both in terms of prediction error and accuracy of the estimation of
the parameters. Again, there is no significant difference between PMF and VG.

\subsection{Effect of the noise}
In this subsection we show the accuracy VG, Lasso and PMF as a function of the
noise~$\hat{\sigma}^2$.  We generate data with $n=100, p=100, p_v=20$ and
$\hat{w}_i=1$ for $20$ randomly chosen components $i$.  We vary
$\hat{\sigma}^2$ in the range $10^{-4}$ to $10$ for for two values of the
correlation strength in the inputs $\zeta=0.5, 0.95$. 

\begin{figure}
\bc
\includegraphics[width=0.8\textwidth]{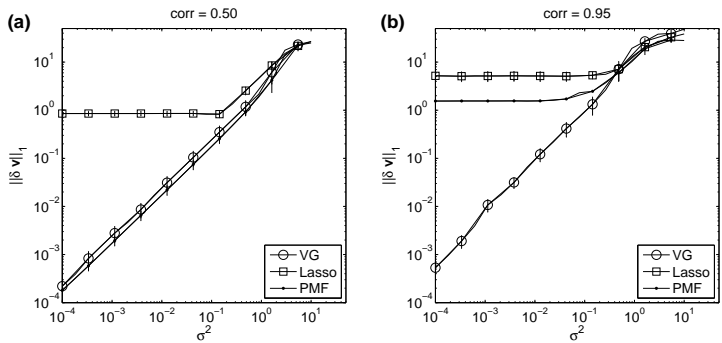}
\ec
\caption{
Accuracy of VG, Lasso and PMF as a function of the noise.  Errorbars of
$\parallel\delta\vv\parallel_1$ for $10$ different random instances.  Data is
generated using $n=100, p=100, p_v=20$ and $\hat{w}_i=1$ for $20$ randomly
chosen components $i$.  We consider
two values of the correlation strength in the inputs: \textbf{(a)} weakly
correlated inputs $\zeta=0.5$ and \textbf{(b)}~strongly correlated inputs
$\zeta=0.95$.
For PMF we choose the best solution (the one with highest value of the bound)
for $10$ different random initializations for each of the $10$ instances.
}
\label{plot_sigma}
\end{figure}

For weakly correlated inputs, Fig.~\ref{plot_sigma}a., we distinguish three
noise domains: for large noise all methods produce errors of $\cO(1)$ and fail
to find the predictive features.  For intermediate and low noise levels, VG and
PMF are significantly better than Lasso.  In the limit of zero noise, the error
of VG and PMF keeps on decreasing whereas the Lasso error saturates to a
constant value. 

For strongly correlated inputs, Fig.~\ref{plot_sigma}b., we observe
that whereas the error of VG scales approximately as before,
PMF gets stuck in local minima in some instances, yielding worse
average performance than VG.  See section~\ref{discussion} for a further
discussion of this point.

\subsection{Analysis of consistency: VG vs Lasso}
\begin{figure}
\bc
\includegraphics[width=0.8\textwidth]{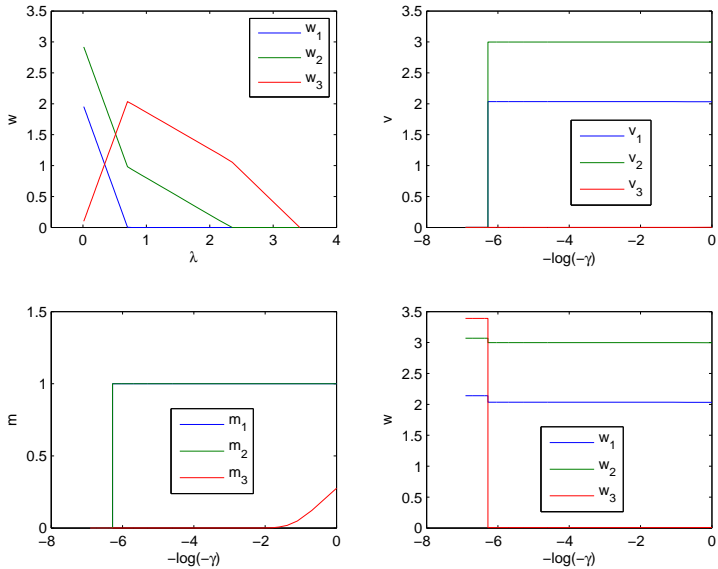}
\ec
\caption{
(Color online) Lasso and VG solution for the inconsistent Example a of
\cite{Zhao:2006:MSC:1248547.1248637}.  Top left: Lasso solution versus
$\lambda$ is called inconsistent because it does not contain a $\lambda$ for
which the correct sparsity ($w_{1,2}\ne 0, w_3=0$) is obtained.
Top right: the VG solution for $\vv$ versus $\gamma$ contains large range of
$\gamma$ for which the correct solution is obtained.  Bottom left: VG
solution for $\vm$ (curves for $m_{1,2}$ are identical). Bottom right: VG
solution for $\vw$. }
\label{zhao1}
\end{figure}

It is well-known that the Lasso method may yield inconsistent results when
input variables are correlated. In \cite{Zhao:2006:MSC:1248547.1248637},
necessary and sufficient conditions for consistency are derived. In addition,
they give a number of examples where Lasso gives inconsistent results.  Their
simplest example has three input variables, $x_1,x_2,x_3$. 
$x_1,x_2,\xi,e$ are independent and Normal distributed random
variables, $x_3=2/3 x_1 + 2/3 x_2 + \xi$ and $y=\sum_{i=1}^3 \hat{w}_i
x_i + e$, $p=1000$. When $\hat{w}=(-2,3,0)$ (Example b) this example is consistent, but
when $\hat{w}=(2,3,0)$ (Example a) this example violates the consistency
condition. The Lasso and VG solution for Example a for different values of $\lambda$ and $\gamma$ are
shown in fig.~\ref{zhao1}a,b, respectively. The VG solution $v_i=m_i w_i$ in terms of
$m_i$ and $w_i$ is shown in fig.~\ref{zhao1}c,d.
The average results over 100 instances for Example~a and Example~b are shown in
table~\ref{zhao2}.
\begin{table}
\bc
\begin{tabular}{l|r|r||r|r}
			&\multicolumn{2}{|c||}{Example a}  &\multicolumn{2}{|c}{Example b} \\
			&$\|\delta\vv\|_1$	&	
$\max(|v_3|)$&$\|\delta\vv\|_1$ &       $\max(|v_3|)$\\\hline
Ridge		&$0.64 \pm 0.18$    &$0.48$	&$0.02 \pm 0.02$&$0.27$\\
Lasso		&$0.19 \pm 0.14$    &$0.30$	&$0.00 \pm 0.00$&$0.00$\\
VG   		&$0.05 \pm 0.03$    &$0.00$	&$0.00 \pm 0.00$&$0.00$
\end{tabular}
\ec
\caption{Accuracy of Ridge, Lasso and VG for Example 1a,b from
\cite{Zhao:2006:MSC:1248547.1248637}. $p=p_v=1000$. Parameters $\lambda$ (Ridge
and Lasso) and $\gamma$ (VG) optimized through cross validation. $\|\delta
\vv\|_{1}$ as before, $\max(|v_3|)$ is maximum over $100$ trials of the absolute
value of $v_3$.  Example a is inconsistent for Lasso and yields much larger
errors than the VG. Example b is consistent and the quality of the Lasso
and VG are similar. Ridge regression is bad for both examples.
}
\label{zhao2}
\end{table}
We see that the VG does not suffer from inconsistency and always finds the
correct solution. This is remarkable as one might have feared that the
non-convexity of the VG would result in sub-optimal local minima. 

\subsection{Boston-housing dataset: VG vs PMF}
We now focus on comparing in more detail the performance of VG with PMF.  In
\cite{titsias2012}, the Boston-housing
dataset\footnote{\texttt{http://archive.ics.uci.edu/ml/datasets/Housing}} is
used to test the accuracy of the PMF approximation.

This is a linear regression problem that consists of $456$ training examples
with one-dimensional response variable $y$ and $13$ predictors that include
housing values.  We use here the same setup as in \cite{titsias2012} to compare
VG with PMF. For PMF, hyperparameters were fixed to values
$\sigma=0.1\times\text{var}(y), \pi = 0.25, \sigma_w = 1$ where var(y) denotes
the output variance. For the VG, we use $\beta = 1/\sigma^2$, $\gamma =
\log(\pi/(1-\pi))$ and hyperparameter $\sigma_w$ is implicitly equal to
$\infty$ in the VG
(see Appendix \ref{sec:pmf} for details of how both models compare).  Since
$\gamma$ and $\beta$ are given, the VG algorithm reduces to iterate
eqs.~\eqref{m} and~\eqref{v} starting from a random $\vm$.  Similarly, the PMF
reduces to perform an E-step given the fixed hyperparameter values.

As in \cite{titsias2012}, we use random initial values for the variational
parameters \emph{between} $0$ and~$1$ (\emph{soft} initialization) and random
values \emph{equal to} $0$ or $1$ (\emph{hard} initialization).  We considered
as ground truth $\hat{w}\equiv\vw^\text{tr}$ the result of the efficient paired Gibbs
sampler developed in \cite{titsias2012}.

\begin{table}
\bc
\begin{tabular}{l|c|c}
 & \emph{soft}-error & \emph{extreme}-error \\\hline
 PMF \cite{titsias2012} & 0.208 [0.002, 0.454]  & 0.204 [0.002, 0.454]  \\
 PMF & 0.237 [0.001, 0.454]  & 0.209 [0.001, 0.454] \\
 VG & 0.006 [0.006, 0.006]  & 0.006 [0.006, 0.006] \\
\end{tabular}
\ec
\caption{
Comparison of VG and PMF in the Boston-housing dataset in terms of
approximating the ground-truth $\hat{w}$.  Average errors 
$\|\delta\vv\|_1=\sum_{i=1}^n |v_i-\hat{w}_i|$
, with $v_i$ the approximation of VG or PMF, together with $95\%$
confidence intervals (given by percentiles) obtained after $300$ 
random initializations for both soft and extreme initializations.
}
\label{housing}
\end{table}
Table \ref{housing} shows the results.  The first and second rows show the
errors reported in \cite{titsias2012} and the errors that we obtain using their
software, respectively.  We observe a small discrepancy in the average errors.
However, if we consider the percentiles, the results are consistent.  In
practice, what we observe is that PMF finds two local optima depending on the
initialization: one is the correct solution (error $\approx 10^{-3}$) whereas
the other has error $0.454$. These two solutions are found
equally often for both soft or hard initializations, showing no dependence on
the type of initialization, in agreement with \cite{titsias2012}.

The results of VG are shown on the third row.  Contrary to PMF, the VG shows no
dependence on the initialization and always finds a solution with an error of
order $10^{-3}$.  These results give evidence that the combined variational/MAP
approach of VG can be better than PMF avoiding local minima.

\subsection{Dependence on the Number of Samples}
We now analyze the performance of all considered methods as a function of the
proportion of samples available. We first analyze the case when inputs are not
correlated and then consider correlations of practical relevance that 
appear in genetic datasets.

For these experiments, we generate the data for dimension $n=500$ and noise
level $\beta=1$.  We explore two scenarios: very sparse problems with only
$10\%$ of active predictors and denser problems with $25\%$ of active
predictors. The weights of the active predictors are set to~$1$.

\subsubsection{Uncorrelated Case}
\begin{figure*}
\bc
\includegraphics[width=.65\textwidth]{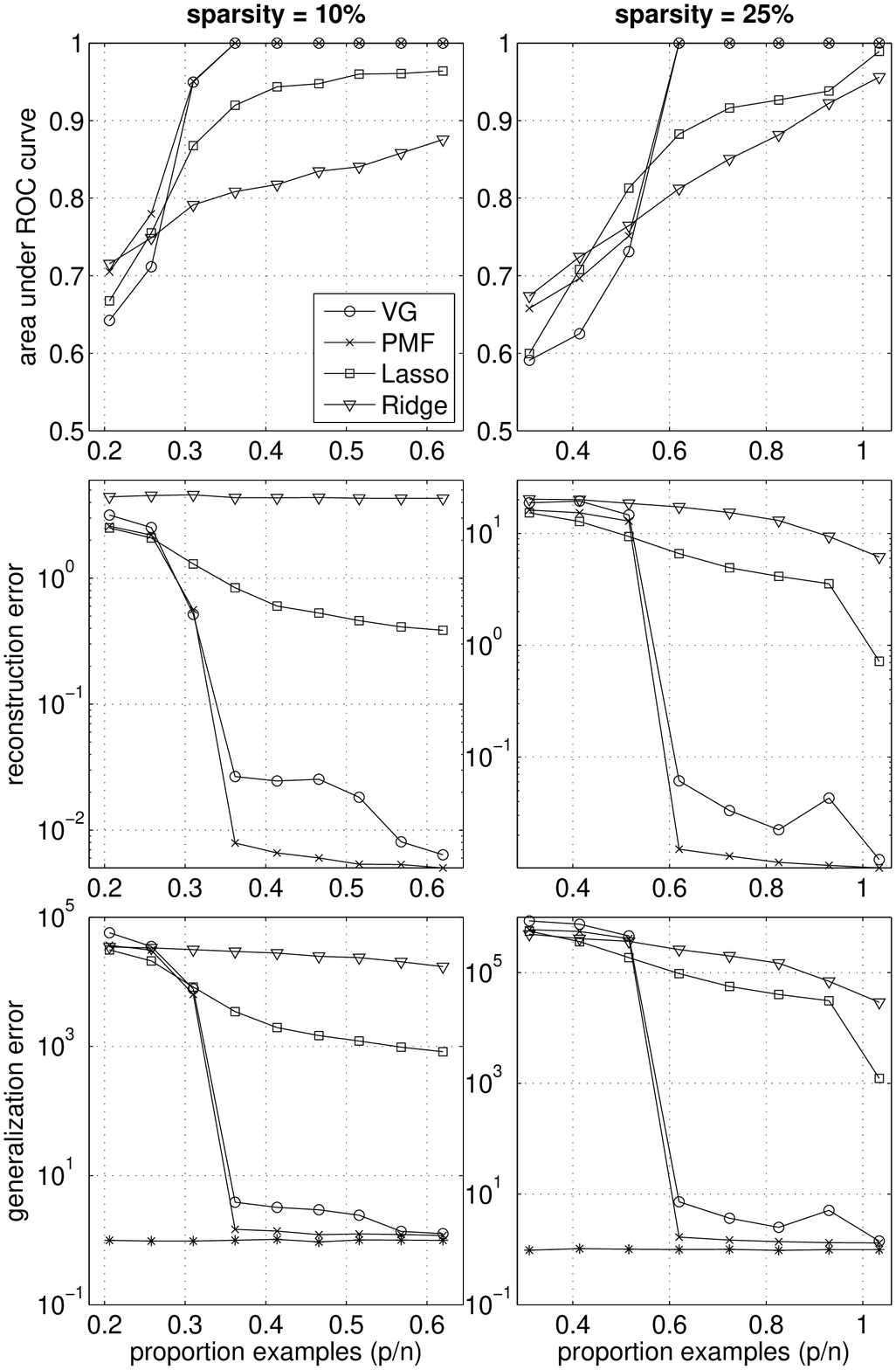}
\ec
\caption{
\textbf{Uncorrelated case}: Performance as a function of
number of training samples $p$ for two levels of sparsity ($10\%$ and
$25\%$ of non-zero entries).  For each value averages over $20$ runs are
plotted.  \textbf{Top:} area under the ROC curves (see text for definition).
\textbf{Middle:}~reconstruction error, defined as
$\|\delta\vv\|_1=\sum_{i=1}^n |v_i-\hat{w}_i|$.
\textbf{Bottom:} generalization
error, defined as the MSE in the test set.
For all methods except for PMF, train set size is $p$
and validation sets size $p_v=p/30$.  For PMF the
training set has size $p + p_v$. Lowest curve shows theoretically optimal generalization 
error obtained by using the target weights from which the data is generated.
}
\label{fig:toy}
\end{figure*}

Figure~\ref{fig:toy} shows results of performance for uncorrelated inputs.  Top
plots show the area under the Receiver Operating Characteristic (ROC) curve.
The ROC curve is calculated by thresholding the weight estimates.  Those
weights that lie above (below) the threshold are considered as active
(inactive) predictors.  
The ROC curve plots the fraction of true positives versus the fraction of false positives for all threshold values.
The area under the curve
measures the ability of the method to correctly classify those predictors that
are and are not active.  A value of~1 for the area represents a perfect classification
whereas 0.5 represents random classification. The ROC is plotted as a function of the
fraction of samples relative to the number of inputs: $p/n$.

For both VG and PMF, we observe in all performance measures a transition from a
regime where solutions are poor to a regime with almost perfect recovery.  This
transition, not noticeable in the other (convex) methods, occurs at around
$35\%$ of examples for $10\%$ of sparsity (left column) and shifts to higher
values for denser problems ($\approx 60\%$ for $25\%$ of sparsity, right
column).  

If we compare VG with PMF we see that in the regime where both methods perform well,
PMF performs slightly better than VG in terms of reconstruction error but
their performance is identical in terms of the area under the ROC curve.  The difference 
between VG and PMF is slightly more
pronounced for denser problems.  We also see that Lasso performs
better than Ridge regression, but the difference between both methods tends to
be smaller for denser problems. Both Lasso and ridge regression are significantly worse than
VG and PMF.

\subsubsection{Correlated case: Genetic dataset}
\begin{figure*}
\bc
\begin{tabular}[t]{cc}
    \includegraphics[width=.33\textwidth]{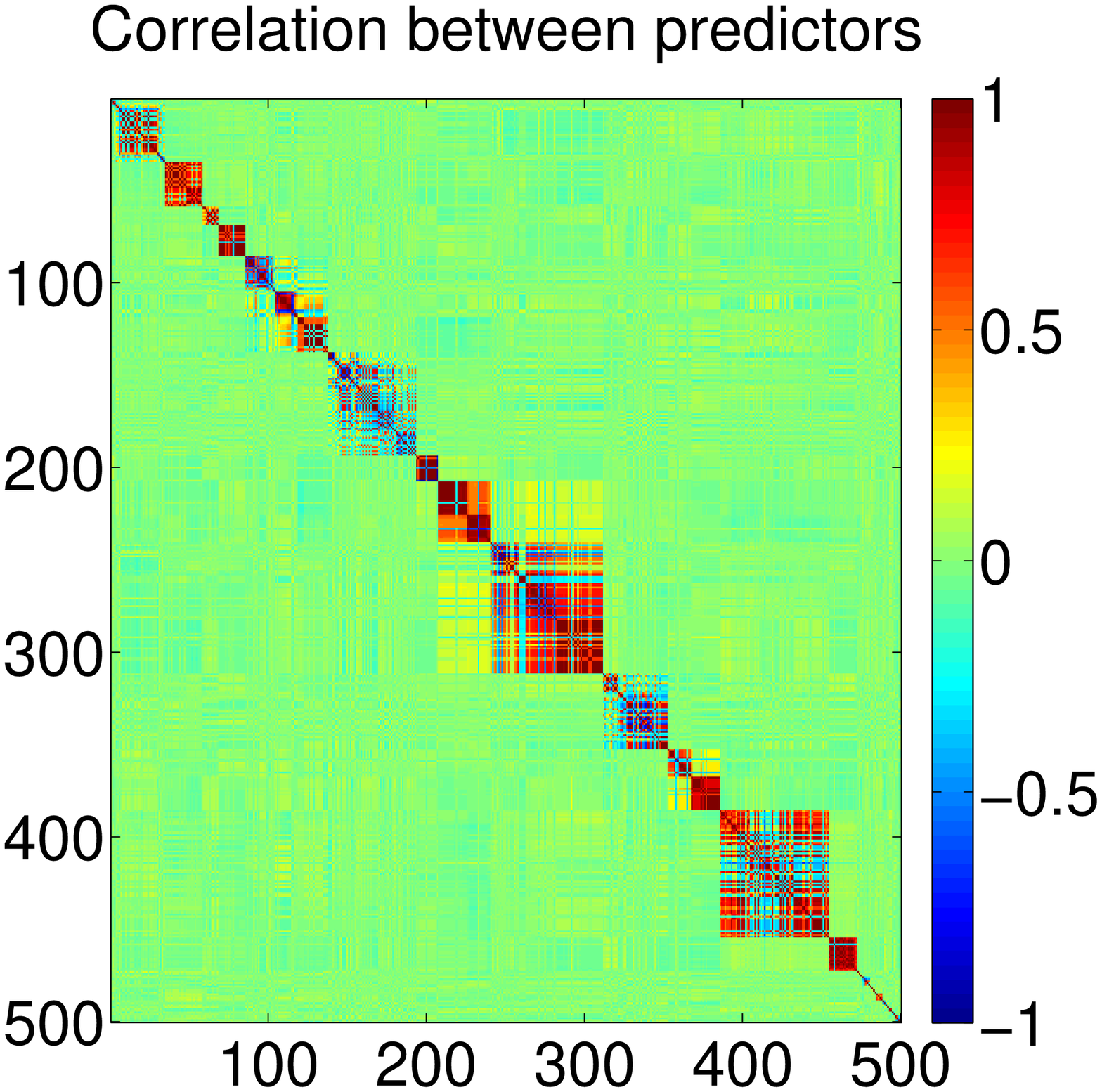} & 
    \includegraphics[width=.65\textwidth]{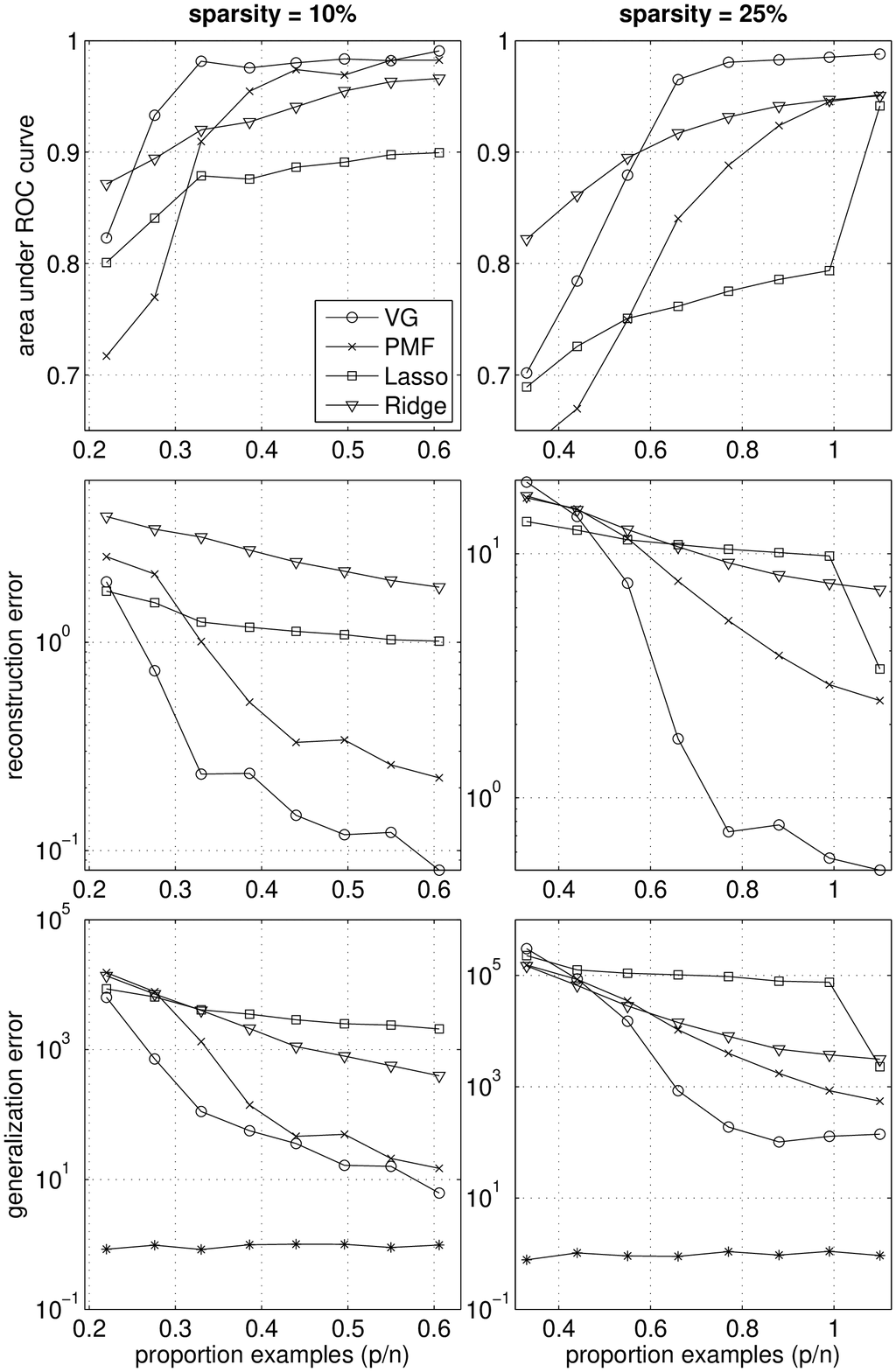} 
    \end{tabular}
\ec
\caption{
\textbf{Correlated case}: \textbf{(LEFT)} (Color online) Example of 
input correlation matrix. \textbf{(RIGHT)} Performance as a function of
number of training samples $p$ for two levels of sparsity ($10\%$ and
$25\%$ of non-zero entries).  For each value averages over $20$ runs are
plotted.  \textbf{Top:} area under the ROC curves (see text for definition).
\textbf{Middle:}~reconstruction error, defined as
$\|\delta\vv\|_1=\sum_{i=1}^n |v_i-\hat{w}_i|$.
\textbf{Bottom:} generalization
error, defined as the MSE in the validation set.
For all methods except for PMF, train set size is $p$
and validation sets size $p_v=p/10$.  For PMF the
training set has size $p + p_v$.
}
\label{fig:correl}
\end{figure*}
We now consider correlated inputs. We use input data obtained from a genetic domain,
where inputs $x_i$ denote single nucleotide polymorphisms (SNPs) that have
values $x_i=\{0,1,2\}$. SNPs typically show correlations structured in blocks,
where nearby SNPs are highly correlated, but show no dependence on distant
SNPs. An example of such correlation matrix can be seen in Figure
\ref{fig:correl} (left). The output data are generated as above.

Figure \ref{fig:correl} (right) shows the results.  Contrary to the
uncorrelated case, the existence of strong correlations between some of the
predictors prevents a clear distinction between solution regimes as a function
of training set size.  We observe, as before, that both VG and PMF are the
preferable methods for sufficiently large training set size. In the three performance measures considered, VG performs better or 
comparable to PMF.  Interestingly, the difference between VG and PMF becomes more
significant for denser problems, when we expect more difficulty due to more
presence of local minima.

\subsection{Scaling with dimension $n$}
\begin{figure*}
\bc
    \includegraphics[width=\textwidth]{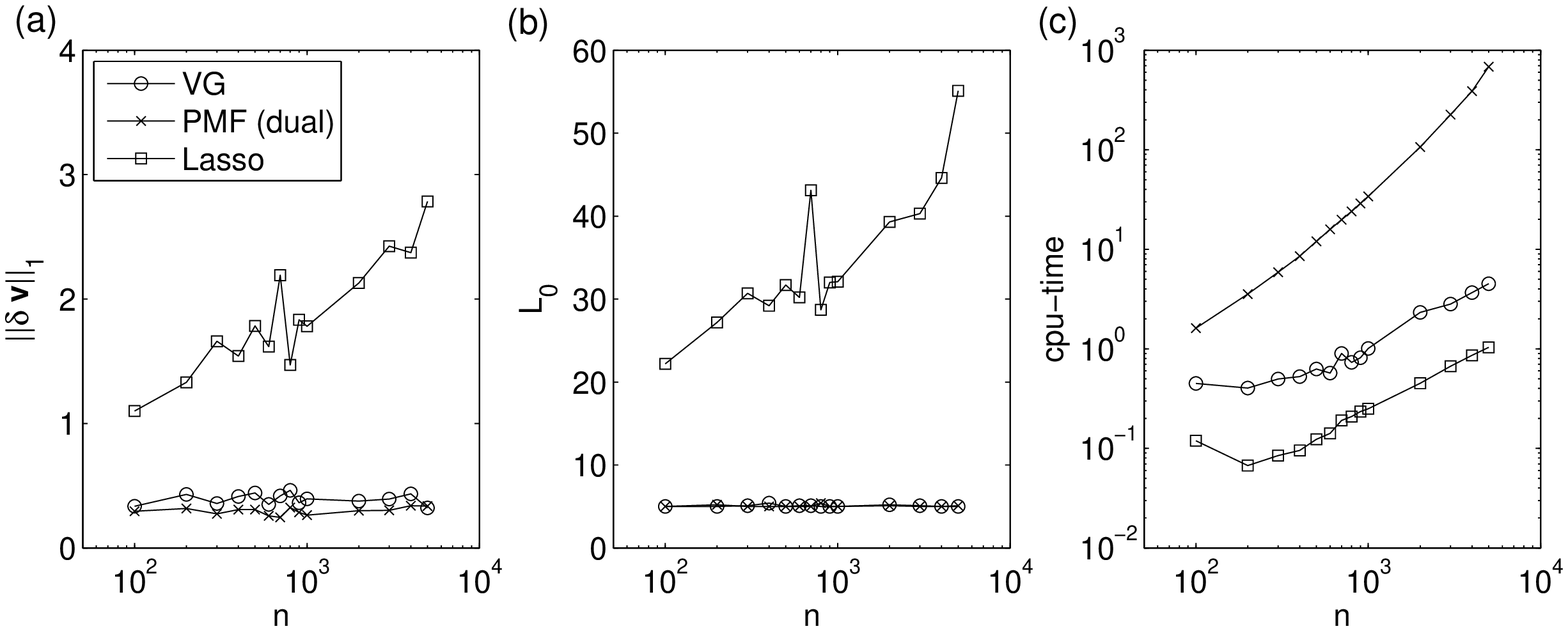} 
\ec
\caption{
\textbf{Scaling with $n$}:
performance of VG, PMF and Lasso as a function of the number of features $n$.
Data are generated as in Example 2. $p=100,p_v=100,\beta=2,\zeta=0$.
}
\label{fig:scaling}
\end{figure*}
We conclude our empirical study by analyzing how the methods scale, both in terms of the quality of the 
solution as in terms of CPU times, as a
function of the number of features $n$ for a constant number of samples.
We use the data as in Example 2 above, with uncorrelated inputs. 

Figure \ref{fig:scaling} shows the results for VG, PMF and Lasso.  For the
VG, we use the dual method described in the appendix \ref{dual}.
Fig.~\ref{fig:scaling}a shows that the VG and PMF have constant quality in
terms of the error $\|\delta\vv\|_1$, whereas the quality of the Lasso
deteriorates with $n$.  Fig.~\ref{fig:scaling}b shows that the VG and PMF
have close to optimal norms $L_0=5$ and that the $L_0$ norm of the Lasso
deteriorates with $n$.  Fig.~\ref{fig:scaling}c shows that the computation time
of all methods scales approximately linear with $n$.  Lasso is significantly
faster than VG and PMF, and VG is significantly faster than PMF. Note, however
that the VG and the PMF methods are implemented in Matlab whereas the Lasso
method uses an optimized Fortran implementation.

%

\section{Discussion}
\label{discussion}
In this paper, we have introduced a new variational method for sparse regression
using $L_0$ penalty. We have presented a minimal version of the
model with no (hierarchical) prior distributions to highlight 
some important features: the variational ridge term that dynamically
regularizes the regression; the input-output behavior as a smoothed
version of hard feature selection; a phase plot that shows when the
variational solution is unique in the orthogonal design case for
different $p,\rho,\gamma$.  We have also shown numerically that the VG
is efficient and accurate and yields results that significantly
outperform other considered methods.

The VG suffers from local minima as can be expected for any method that needs
to solve a non-convex problem, like the PMF.  From the numerical experiments we
can conclude that VG is on average preferable to PMF in practical scenarios
with strongly correlated inputs and/or moderately sparse problems, where the
local minima problem is more severe.  Although we have no principled solution
for the local minima problem, we think that the combined variational/MAP
approach together with the the annealing procedure that results from increasing
$\gamma$, followed by a "heating" phase to detect hysteresis works well in
practice, helping to avoid local minima.  Another obvious approach is to use
multiple restarts or using more powerful approximations, such as structured
mean field approximation or belief propagation.  We remark that the PMF in the
general setting \cite{titsias2012} includes an extra layer of flexibility that
can be used to capture correlations between input variables.  Such extensions
can also be considered for VG.



We have not explored the use of different priors on $\vw$ or on
$\beta$. In addition, a prior could be imposed on $\gamma$.  It is
likely that for particular problems the use of a suitable prior
could further improve the results.

We have seen that the performance of the VG is excellent in the zero noise
limit. In this limit, the regression problem reduces to a compressed sensing
problem \cite{candes2005,donoho2006}. The performance of compressed sensing
with $L_q$ sparseness penalty was analyzed theoretically in
\cite{kabashima2009}, showing the superiority of the $L_1$ penalty in
comparison to the $L_2$ penalty and suggesting the optimality of the $L_0$
penalty. Our numerical result are in agreement with this finding.

Our implementation uses parallel updating of Eqs.~\ref{m}-\ref{beta}, or
Eqs.~\ref{m},\ref{w2},~\ref{A}-\ref{lambda1} when using the dual
formulation. One may consider also a sequential updating. This was done
successfully for the Lasso based on the idea of the Gauss-Seidel algorithm 
\cite{friedman2009}. The advantage of such an approach is that each update
is linear in both $n$ and $p$, since only the non-zero components need to
be updated. However, the number of updates to converge will be larger. The
proof of convergence for such a coordinate descend method for the VG is
likely to be more complex than for the Lasso due to non-convexity. As a
result, a smoothing parameter $\eta \ne 1$ (see Algorithm~\ref{algorithm}) 
may still be required.


\subsubsection*{Acknowledgments}
We would like to thank M. Titsias for providing the code of PMF and specially
the Boston Housing files.  We also thank Kevin Sharp and Wim Wiegenrick for
useful discussions and anonymous reviewers for helping on improving the
manuscript.

 \bibliography{1}

\bibliographystyle{unsrt} 

\appendix

\section{: Phase plot computation for the orthogonal case}
\label{appendix:univ}
\begin{figure}[!h]
\bc
\includegraphics[width=0.75\textwidth]{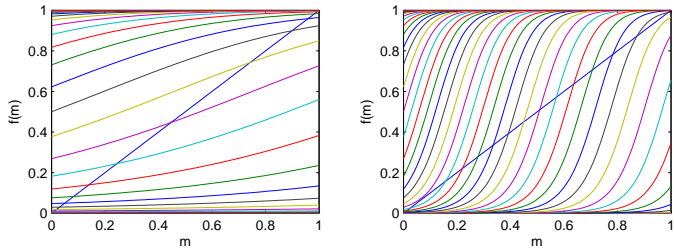}
\ec
\caption{(Color online) $f(m)$ vs $m$. 
Left (a): $p=100,\gamma=-10$, different lines correspond to
different values of $0<\rho<1$ (higher lines are higher $\rho$). The solution for $m$ is given by the intersection
$f$ with the diagonal line. The solution for $m$ is unique and increases with
increasing $\rho$.
Right (b): Same as left, but with $p=100, \gamma=-30$. Depending on $\rho$,
there are one or three
solutions for $m$. The solutions close to $m\approx 0,1$ correspond to local minima
of $F$. The intermediate solution corresponds to a local maximum of $F$.}
\label{file4b}
\end{figure}

In the uni-variate case, $f(m)$ in Eq.~\ref{univ1} is
an increasing function of $m$ and crosses the line $m$ either 1 or
three times, depending on the values of $p$ and $\gamma$ (see fig.~\ref{file4b}).
In the multivariate orthogonal case, this is still true, since 
the influence of other features is only
through $\beta$. We can thus write $\beta^{-1}=\sigma_y^2(1-\rho m
-\delta)$, where $0\le \delta<1$ is a function of the variational parameters
of the other features.
Thus, there are regions of parameter space $\gamma,p,\rho$ where
the uni-variate solution is unique and others for which there are
two stable solutions.

The transition between these two regions is when $f'(m)=1$
and $f(m)=m$. These two equations imply
\bea
\left(1+\frac{p}{2}\right)\rho^2m^2-\left(2\rho(1-\delta)
+\frac{p}{2}\rho^2\right)m+(1-\delta)^2=0
\label{apendix_univ1}
\eea
This quadratic equation in $m$ has either zero, one or two solutions, corresponding
to no touching, touching once and touching twice, respectively.
Denote $a=(1+\frac{p}{2})\rho^2, b=2\rho(1-\delta)+\frac{p}{2}\rho^2$.
The critical value for
$\rho,p$ is when Eq.~\ref{apendix_univ1} has one solution for $m$,
which occurs when 
\bea
D&=&b^2-4 a (1-\delta)^2=
\frac{p}{2}\rho^2\left(\rho-\rho^*\right)\left(\frac{p}{2}\rho+2(1-\delta)+2(1-\delta)\sqrt{1+\frac{p}{2}}\right)=0\nonumber\\
\rho^*&=&\frac{4}{p}(1-\delta)\left(\sqrt{1+\frac{p}{2}}-1\right)
\label{rhostar}
\eea
Thus, $D$ is positive when $\rho>\rho^*$ and Eq.~\ref{apendix_univ1} has two
solutions for $m$. We denote these solutions by 
$m_{1,2}=\frac{b\pm \sqrt{D}}{2a}$. 
Note, that the solutions in these critical
points only depend on
$\rho,p$. For each of these solutions we must find a $\gamma$ such that $f(m)=m$,
which is given by
\bea
\gamma_i&=&\log \frac{m_i}{1-m_i}-\frac{p}{2}\frac{\rho}{1-\rho m_i}\qquad i=1,2
\label{appendix_univ2}
\eea
It is easy to see that the smallest of these solutions $m_1<m_2$ corresponds to a
local maximum of the free energy and can be discarded. Thus, when
$\rho>\rho^*$ and $\gamma_2<\gamma<\gamma_1$ two stable variational
solutions $m\approx 0,1$ co-exist.

When $\rho<\rho^*$, Eq.~\ref{apendix_univ1} has no solutions for
$m$.  In this case the conditions $f'(m)=1$ and $f(m)=m$ cannot be
jointly satisfied and the variational solution is unique.

From Eq.~\ref{rhostar} we see that $\rho^*$ is a decreasing function of $p$ and when $p\gg 1$, $\rho^*\approx
2\sqrt{\frac{2}{p}}$.
In the critical point, where $\rho=\rho^*(p)$,
$m=b/2a\approx\frac{1}{2}\left(1+\sqrt{\frac{2}{p}}\right)$ and 
\bea
\gamma^*\approx -\sqrt{2
p}(1-\delta)
\label{gammastar}
\eea
When $\rho<\rho^*$ {\em or} $\gamma>\gamma^*$ the variational solution is
unique.
We illustrate the phase plot $\rho,\gamma$ for $p=100$ in fig.~\ref{file4c}a.

\section{: Dual Formulation}
\label{dual}
The solution of the system of Eqs.~\ref{m}-\ref{beta} by fixed point
iteration requires the repeated solution of the $n$ dimensional linear
system $\chi' \vw = \vb$. When $n> p$, we can obtain a more
efficient method using a dual formulation.

We define new variables $z^\mu=\sum_i m_i w_i x_i^\mu$ and add
Lagrange multipliers $\lambda^\mu$:
\bea
F&=&
-\frac{p}{2}\log\frac{\beta}{2\pi}+\frac{\beta}{2}\sum_{\mu}^p (z^\mu-y^\mu)^2
+\frac{\beta p}{2}\sum_i m_i(1-m_i)w_i^2 \chi_{ii}\nonumber\\
&-&\gamma\sum_{i=1}^n m_i 
+\sum_{i=1}^n \left(m_i \log m_i+(1-m_i)\log (1-m_i)\right)\nonumber\\
&+&\sum_\mu \lambda^\mu( z^\mu-\sum_i m_i w_i x_i^\mu)
\label{F2}
\eea

We compute the derivatives of Eq.~\ref{F2}:
\beaa
\frac{\partial F}{\partial w_i}&=&m_i\left(\beta p
(1-m_i)\chi_{ii}w_i-\sum_\mu \lambda^\mu x_i^\mu\right)\\
\frac{\partial F}{\partial z^\mu}&=&\beta (z^\mu-y^\mu)+\lambda^\mu\\
\frac{\partial F}{\partial \beta}&=&-\frac{p}{2\beta} + \frac{1}{2}\sum_{\mu}^p (z^\mu-y^\mu)^2+\frac{p}{2}\sum_i
m_i(1-m_i)w_i^2 \chi_{ii}\\
\frac{\partial F}{\partial m_i}&=&\frac{\beta p}{2}(1-2
m_i)w_i^2\chi_{ii}-\gamma +\sigma^{-1}(m_i)-\sum_\mu \lambda_\mu w_i
x_i^\mu\\
\frac{\partial F}{\partial \lambda^\mu}&=& z^\mu -\sum_i m_i w_i
x_i^\mu
\eeaa

By setting $\frac{\partial F}{\partial w_i}=\frac{\partial F}{\partial
z^\mu}=0$ we obtain 
\bea
w_i&=&\frac{1}{\beta p \chi_{ii}}\frac{1}{1-m_i}\sum_\mu \lambda^\mu
x_i^\mu\label{w2}
\eea
and $z^\mu=y^\mu -\frac{1}{\beta}\lambda^\mu$.
Setting the remaining derivatives to zero, and eliminating $w_i$ and
$z^\mu$ we 
obtain Eq.~\ref{m} and
\bea
\beta&=&\frac{1}{p}\sum_{\mu\nu}\lambda_\mu \lambda_\nu A_{\mu\nu}
\label{beta2}\\
\beta y^\mu&=& \sum_\nu A_{\mu\nu}\lambda^\nu \label{lambda}
\eea
with $A_{\mu\nu}$ given by 
\bea
A_{\mu\nu}&=&\delta_{\mu\nu}+\frac{1}{p}\sum_i
\frac{m_i}{1-m_i}\frac{x_i^\mu x_i^\nu}{\chi_{ii}}\label{A}
\eea
For given $A_{\mu\nu}$, let $\hat{y}$ denote the solution of
\bea
\sum_{\nu=1}^p A_{\mu\nu} \hat{y}^\nu&=& y^\mu\label{yhat}
\eea
Then it is easy to verify that
\bea
\frac{1}{\beta}&=&\frac{1}{p}\sum_\mu \hat{y}^\mu y^\mu\label{beta1}\\
\lambda^\mu&=&\beta \hat{y}^\mu\label{lambda1}
\eea
solve the system of Eqs.~\ref{beta2}-\ref{lambda}.

\section{: Relation with the Paired-Mean Field approximation}
\label{sec:pmf}
The VG shares many similarities with the recently proposed paired mean field
(PMF) variational approach \cite{titsias2012}.  Here we relate
both approaches in terms of three different aspects: the probabilistic model,
the variational approximation and the optimization algorithm.
\begin{description}
\item[Model]: 
The model considered for the PMF variational approximation is defined for
multiple outputs and considers a linear combination of basis functions governed
by a Gaussian process.  To relate this model to the one presented in this work,
we consider the one-dimensional output without the extra input layer. 

The spike and slab model \cite{george1993} considers a linear regression model
of the form:
\begin{align*}
y &= \sum_{i=0}^{n}\hat{v}_ix_i + \xi\\
\hat{v}_i &\sim \pi\mathcal{N}(\hat{v}_i|0,\sigma^2_w) + (1-\pi)\delta_0(\hat{v}_i),
\qquad \forall i.  
\end{align*}
That is, the prior over the weights is factorized, with each weight distributed
according to a mixture distribution: with probability $\pi$, each $\hat{v}_i$
is drawn from a Gaussian centered at zero with variance $\sigma^2_w$, and with
probability $1-\pi$, each $\hat{v}_i$ is zero.  The sparsity of the solution is
controlled by $\pi$, either directly or by specifying a prior over $\pi$.

Observe that we can equivalently write $\hat{v}_i$ as the product of a
Bernoulli random variable $s_i \sim \pi^{s_i} (1-\pi)^{1-s_i}$ and a Gaussian
random variable $w_i \sim \mathcal{N}(w_i|0,\sigma_w^2)$, which is the
reparameterization used in \cite{titsias2012}.

%
To relate the model used in the VG defined by Eqs.~\eqref{vg} and \eqref{prior}
to the previous one we make the following identifications:
\begin{itemize}
\item
The prior on $w_i$ is flat, which corresponds to setting $\sigma_w = \infty$ in
\cite{titsias2012}.
\item 
$\gamma = \log(\pi/(1-\pi))$.
\end{itemize}
Thus, the spike and slab model \cite{george1993} and the model considered by \cite{titsias2012}
are identical and both models are identical to the model considered in this
paper when a Gaussian prior is placed over the weights.

\item[Variational approximation]:
The PMF variational distribution places each weight $w_i$ and
bit $s_i$ in the same factor:
\begin{align}
\label{eq:pmf}
q(\vec{w},\vec{s})&=\prod_{i=1}^nq_i(w_i,s_i).
\end{align}
On the contrary, the VG reduces to the classical factorized variational
approach under the restriction that the posterior for the weight is a delta
function.

\item[Algorithm]:
The optimization in \cite{titsias2012} uses an EM algorithm that alternates
between expected values of the latent variables $\vw,\vs$ (E-Step) and
optimization of hyperparameters $\{\sigma_y^2,\sigma^2_w,\pi\}$ (M-Step).

The VG method differs mainly in two points. The VG method:
\begin{itemize}
\item Computes expectation of $\vs$ (denoted by $\vm$) 
but finds MAP solution for $\vw$.
\item 
Searches the space of solutions using a forward and a backward
sequential search over hyperparameter $\gamma$ using a validation set.  For a
given $\gamma$, the rest of the parameters are optimized using a training set
and initialized with a 'warm' solution from the previous step (see
Algorithm~\ref{algorithm}).
\end{itemize}


%
\end{description}

\end{document}